\documentclass[sn-mathphys,Numbered]{sn-jnl}
\usepackage{graphicx}
\usepackage{multirow}
\usepackage{amsmath,amssymb,amsfonts}
\usepackage{amsthm}
\usepackage{mathrsfs}
\usepackage[title]{appendix}
\usepackage{xcolor}
\usepackage{textcomp}
\usepackage{manyfoot}
\usepackage{booktabs}
\usepackage{algorithm}
\usepackage{algorithmicx}
\usepackage{algpseudocode}
\usepackage{listings}
\DeclareUnicodeCharacter{0308}{\"}
\DeclareUnicodeCharacter{030C}{\v{ }}
\DeclareUnicodeCharacter{0301}{\'{ }}
\usepackage[]{units}

\usepackage{xparse}
\usepackage{commath}
\usepackage{longtable}
\usepackage{MnSymbol,wasysym,stmaryrd,ifsym,rotating,marvosym}
\usepackage{euscript,tabularx}
\usepackage{pgfplots}
\usetikzlibrary{pgfplots.groupplots}
\pgfplotsset{compat=1.3}
\usepackage[T1]{fontenc}
\usepackage{textcomp}
\usepackage{pgfplots}
\usepackage{amsmath}
\usepackage{etoolbox}
\pgfplotsset{compat=newest} 
\pgfplotsset{plot coordinates/math parser=false}
\usepackage{graphicx}
\usepackage{dcolumn}

\usepackage{titlesec}
\titleformat{\section}
  {\normalfont\large\bfseries\raggedright}{\thesection.}{0.5em}{}
\titlespacing*{\section}
  {0pt}{3.5ex plus 1ex minus .2ex}{1.5ex plus .2ex}

\titleformat{\subsection}
  {\normalfont\normalsize\bfseries\raggedright}{\thesubsection.}{0.5em}{}
\titlespacing*{\subsection}
  {0pt}{2.5ex plus 1ex minus .2ex}{1ex plus .2ex}

\begin{document}
\title[Article Title]{Diverse dynamics in interacting vortices systems through tunable conservative and non-conservative coupling strengths}

\author*[1,8]{\fnm{Alexandre Abbass} \sur{Hamadeh}}\email{hamadeh@rptu.de or alexandre.hamadeh@universite-paris-saclay.fr}
\author[1]{\fnm{Abbas} \sur{Koujok}}
\author[2]{\fnm{Davi R.} \sur{Rodrigues}}
\author[3]{\fnm{Alejandro} \sur{Riveros}}
\author[4]{\fnm{Vitaliy} \sur{Lomakin}}
\author[5]{\fnm{Giovanni} \sur{Finocchio}}
\author[6]{\fnm{Grégoire} \sur{de Loubens}}
\author[7]{\fnm{Olivier} \sur{Klein}}
\author[1]{\fnm{Philipp} \sur{Pirro}}

\affil*[1]{\orgdiv{Fachbereich Physik and Landesforschungszentrum OPTIMAS}, \orgname{Rheinland-Pf\"alzische Technische Universit\"at Kaiserslautern-Landau}, \orgaddress{\street{} \postcode{67663}, \city{Kaiserslautern}, \state{} \country{Germany}}}

\affil[2]{\orgdiv{Department of Electrical and Information Engineering}, \orgname{Politecnico di Bari}, \orgaddress{\street{} \postcode{70126}, \city{Bari}, \state{} \country{Italy}}}

\affil[3]{\orgdiv{Centro de Investigación en Ingeniería de Materiales, FINARQ}, \orgname{Universidad Central de Chile}, \orgaddress{\street{} \postcode{8330601}, \city{Santiago}, \state{} \country{Chile}}}

\affil[4]{\orgdiv{Center for Memory and Recording Research and Department of Electrical and Computer Engineering}, \orgname{University of California}, \orgaddress{\street{} \postcode{92093-0407}, \city{San Diego, La Jolla}, \state{California} \country{USA}}}

\affil[5]{\orgdiv{Department of Mathematical and Computer Sciences, Physical Sciences and Earth Sciences}, \orgname{University of Messina}, \orgaddress{\street{} \postcode{I -98166}, \city{Messina}, \state{} \country{Italy}}}

\affil[6]{\orgdiv{SPEC, CEA, CNRS}, \orgname{Universit\'e Paris-Saclay}, \orgaddress{\street{} \postcode{91191}, \city{Gif-sur-Yvette}, \state{} \country{France}}}

\affil[7]{\orgdiv{Univ. Grenoble Alpes, CEA, CNRS}, \orgname{Grenoble INP, Spintec}, \orgaddress{\street{} \postcode{38054}, \city{Grenoble}, \state{} \country{France}}}
\affil[8]{\orgdiv{Université Paris-Saclay,}, \orgname{ Centre de Nanosciences et de Nanotechnologies,CNRS}, \orgaddress{\street{} \postcode{91120}, \city{Palaiseau}, \state{} \country{France}}}

\abstract{Magnetic vortices are highly tunable, nonlinear systems with ideal properties for being applied in spin wave emission, data storage, and neuromorphic computing. However, their technological application is impaired by a limited understanding of non-conservative forces, that results in the open challenge of attaining precise control over vortex dynamics in coupled vortex systems. Here, we present an analytical model for the gyrotropic dynamics of coupled magnetic vortices within nano-pillar structures, revealing how conservative and non-conservative forces dictate their complex behavior. Validated by micromagnetic simulations, our model accurately predicts dynamic states, controllable through external current and magnetic field adjustments. The experimental verification in a fabricated nano-pillar device aligns with our predictions, and it showcases the system’s adaptability in dynamical coupling. The unique dynamical states, combined with the system’s tunability and inherent memory, make it an exemplary foundation for reservoir computing. This positions our discovery at the forefront of utilizing magnetic vortex dynamics for innovative computing solutions, marking a leap towards efficient data processing technologies.}


\maketitle

\section{Introduction}\label{sec1}

From macroscopic systems governed by classical mechanics all the way to microscopic systems with dynamics in the quantum realm, coupled systems are integrated in natural phenomena. The mutual interaction and non-linearity inherent to such systems often give rise to dynamics of which diversity is contingent upon the complexity of the systems under study. The interaction of conservative and non-conservative coupling mechanisms with nonlinearity gives rise to particularly rich dynamics. For example, ecological out-of-equilibrium systems and the nervous system exhibit complex coupled dynamics that are essential for their functionality\cite{blasius1999complex,brown2001complex,schlueter2012new,hein2020information,hua2004neural,yang2016generalized,de2018dynamic}. The increasing prevalence of bio-inspired computing applications has led to a surge in interest in such systems, as evidenced by a growing body of literature on the subject \cite{squartini2013advances,datta2014neuro,volos2015memristor,nakada2016pulse,markovic2020physics,csaba2020coupled,krause2021robust}. Nevertheless, the development of scalable systems with tunable conservative and non-conservative couplings, which are crucial for the creation of bio-inspired hardware, represents a significant challenge.

Recent advances in the field of magnetic vortices have demonstrated their potential as highly tunable, nonlinear systems. The formation of magnetic vortices is contingent upon the specific geometries and aspect ratios of the materials in question. The dynamics of these vortices are governed by the interplay of various magnetic interactions \cite{guslienko2004vortex}. The gyrotropic mode, the lowest-frequency excitation of magnetic vortices, has attracted interest for its potential applications in spin wave emission \cite{wintz2016magnetic,mayr2021spin,hamadeh2022hybrid}, data storage \cite{yamada2007electrical,guslienko2008dynamic}, and neuromorphic computing \cite{torrejon2017neuromorphic,shreya2023granular}.  Notwithstanding these advances, significant challenges remain in attaining precise control over vortex dynamics, particularly in coupled vortex systems. A comprehensive understanding of the interactions between these entities, particularly in the presence of non-conservative forces such as spin-polarized currents, is essential for the full exploitation of their potential.
The objective of this study is to address the aforementioned challenges by developing and analyzing a system of coupled magnetic vortices within a nanopillar stack geometry. By precisely tuning conservative and non-conservative interactions, our system offers a promising platform for the next generation of bio-inspired computing technologies. We present an analytical model for the interaction mechanisms of coupled vortex structures, validated through finite-element micromagnetic simulations. The experimental results demonstrate the influence of external magnetic bias and direct currents on the dynamics of vortex cores, thereby elucidating a complex phase space of orbit states. Not only do these findings advance our fundamental understanding of the subject, but they also highlight the potential of incorporating coupled magnetic vortices in a number of applications including reservoir computing and beyond.

\section{Results}\label{sec2}

\subsection{Analytical Model for coupled vortices}

\begin{figure}[!ht]
  \centering
  \includegraphics[width=\columnwidth]{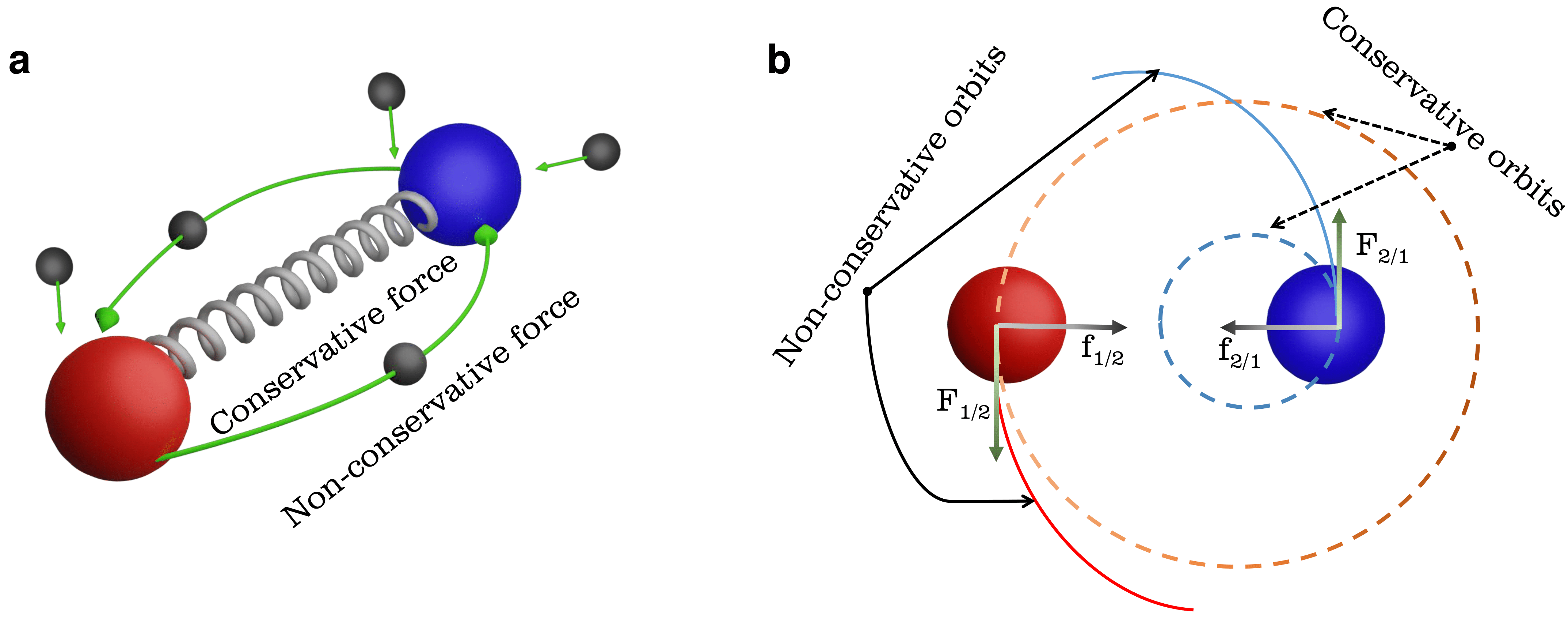}
 \caption{\textbf{Simplified illustration of the conservative and non-conservative interactions between two particle-like objects: }(a) Schematic illustration of the nature of coupling between two particle-like objects (red and blue). The spring between them represents the conservative coupling inherent in their mutual interaction, which persists even in the absence of external excitation. The green arrows symbolize non-conservative forces resulting from interaction with a dynamical particle bath. The scattering of particles in the bath between the two objects generates an additional coupling force. (b) Schematic representation depicting orbits generated by distinct coupling terms. Conservative coupling (f$_{1/2}$ and f$_{2/1}$) predominantly yields closed orbits, driven by energy conservation principles, while non-conservative coupling (F$_{1/2}$ and F$_{2/1}$) fosters a varied orbit unconstrained by energy conservation, adding diversity to the dynamic system. Arrows are represented schematically and their sizes do not represent the magnitude of the vectors.}
  \label{fig:1}
\end{figure}

Vortex Cores (VCs), regardless of their underlying physical system, exhibit dynamic behavior in which their positions along two perpendicular axes within the plane of rotation exhibit a canonical conjugate relationship. Specifically, their positions, denoted by $X$ and $Y$, obey the canonical Poisson bracket rule $\{X,Y\} = \textrm{constant}$, where the constant typically depends on the underlying physical system and is typically linked to a topological quantity. This Poisson bracket is at the root of the Magnus effect that affects the dynamics of rotating objects\cite{swanson1961magnus,dooghin1992optical,bliokh2004topological}. Using this Poisson bracket and adopting the Hamiltonian formalism, we can derive a general system of first-order differential equations of motion to describe the dynamics of coupled rotating particles,
\begin{align}\label{eq:EOMXY}
    G_{i}\dot{X}_{i} &= +\frac{\partial E_{0\,i}}{\partial Y_{i}} + D_{X\,i}\dot{Y}_{i} + \sum_{j}\left(+\frac{\partial E_{ij}}{\partial Y_{i}} + F_{X\,ij}\right), \notag\\ 
    G_{i}\dot{Y}_{i} &= -\frac{\partial E_{0\,i}}{\partial X_{i}} - D_{Y\,i}\dot{X}_{i} +\sum_{j}\left(- \frac{\partial E_{ij}}{\partial X_{i}} + F_{Y\,ij}\right)
\end{align}
where $X_{i}$ and $Y_{i}$ are the coordinates of the center of the rotating particles $i=1,2,\cdots N$ and $G_{i}$ is given by $\{X_{i},Y_{i}\} = 1/G_{i}$ and is often associated to the angular momentum of the underlying physical system. Eqs.\eqref{eq:EOMXY} are versatile and not limited to a single physical system. They can describe the dynamics of magnetic vortices \cite{guslienko2001field}, magnetic skyrmions \cite{mckeever2019characterizing}, and other phenomena across diverse fields, including fluid dynamics \cite{morrison1998hamiltonian}. The terms on the right-hand side represent the different forces acting on each particle: the first two terms describe interactions between each particle and the surrounding medium, associated with constraining potentials ($E_{0}$) and viscous forces ($\propto D$). The last two terms describe interactions between the particle $i$ with the other particles $j$. With the last term $F$ within the sum we introduce the concept of non-conservative coupling between the particles $1$ and $2$. Fig. \ref{fig:1} (a) is a schematic illustrating the conservative and non-conservative coupling mechanisms of the two particles (blue/red spheres). The intrinsic interaction, governed by the mutual force exerted by one object on the other, shown as a gray spring between them, represents the conservative coupling. This terms alone leads to  well known dynamics, e.g. planetary motion. Very important within the context of our study is the  non-conservative force illustrated by green arrows. They represent the interaction with an external particle bath (energy source) whose action depends on the state of both particles ($F=F_{ij}$).

In general, non-conservative coupling arises from interactions between particles and a dynamic particle bath. The magnitude of this coupling is related to the energies and momenta of the particles in the bath. This concept extends its relevance to various disciplines of physics such as fluid dynamics \cite{saffman1995vortex}, optics \cite{dooghin1992optical,bliokh2004topological}, astrophysics \cite{nezlin2013rossby}, high-energy physics \cite{hanany2003vortices}, and magnetism \cite{gobel2021beyond}. It envisions scenarios where interactions between two bodies occur within a dynamics particle cloud. Moreover, it's necessary to note an important distinction: while the conservative forces are bound by the underlying symmetries of the many-body system, non-conservative forces are not bound by these constraints. They can, for example, violate the conservation of total energy and momentum of the considered system. Thus, we emphasize that while conservative coupling cannot generate motion in a damped system such that the system remains in the lowest energy state, non-conservative coupling generates motion that drives the system out of equilibrium. This distinction grants new degrees of freedom in motion, opening up avenues for diverse behaviors. Fig. \ref{fig:1} (b) depicts a common scenario: conservative forces (f$_{1/2}$ and f$_{2/1}$) align with centripetal forces, while non-conservative interactions can generate tangential forces (F$_{1/2}$ and F$_{2/1}$). 

\begin{figure}[!ht]
  \includegraphics[width=\columnwidth]{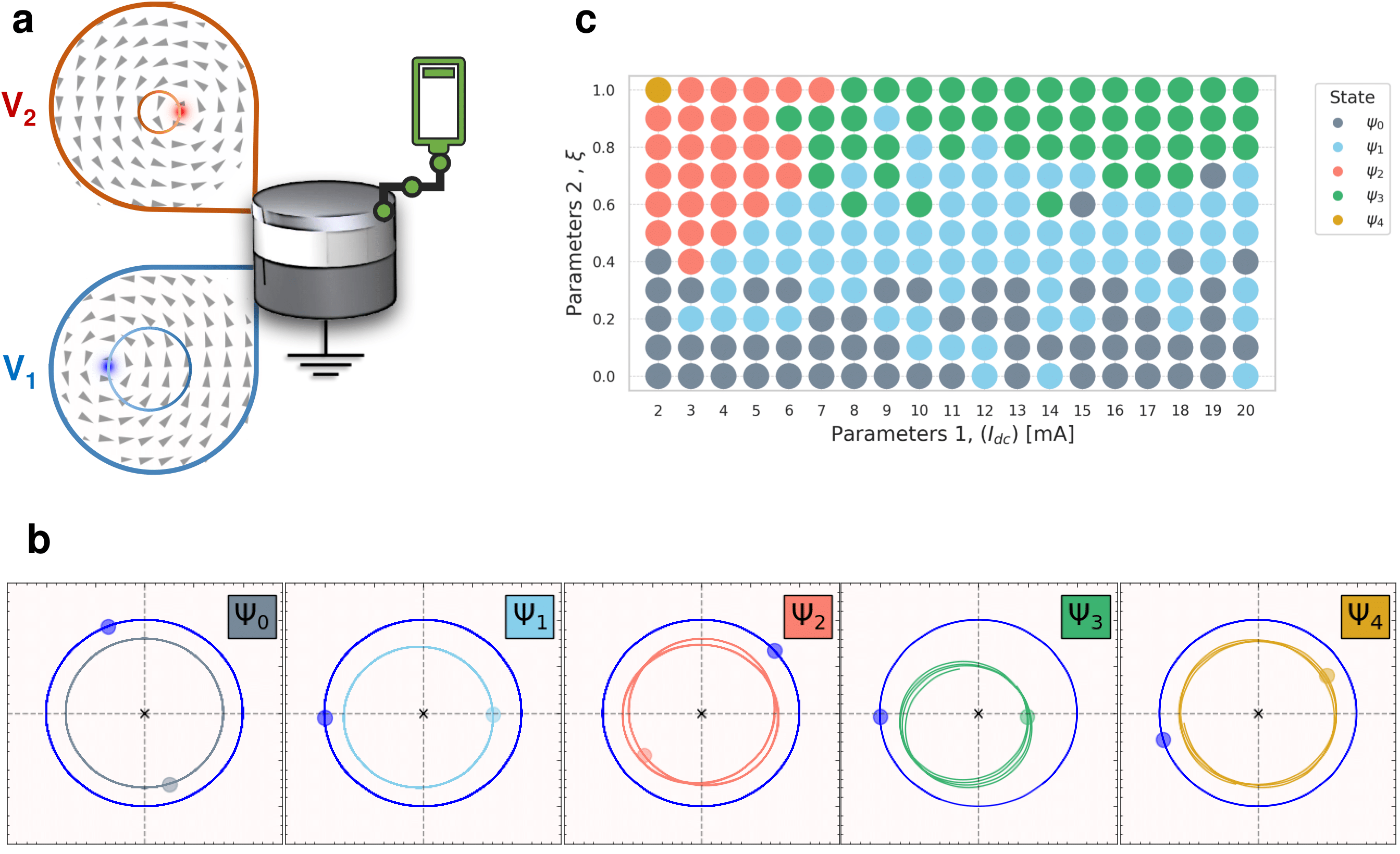}
  \centering
 \caption{\textbf{Analytical model calculations:} (a) Simplified schematic of the studied tri-layered system with coupled vortices V$_{1}$ (bottom layer) and V$_{2}$ (top layer). V$_{1}$ has a negative polarity $\Pi _{1} ^-$. V$_{2}$ has a positive polarity $\Pi _{2} ^+$. (b) Trajectories underwent by the layer 1's (blue) and layer 2's (colored) VCs for states $\Psi_{i}$ (i = 0, 1, 2, 3, 4) as extracted from the analytical calculations. Labels and axis units were removed as the values representing the x and y axes have been normalized. Layer 2's VC trajectories are magnified twice for presentation purposes. (c) Phase diagram of the coupled dynamics, as a function of the control parameters $I_{\mathrm{dc}}$ and $\xi$, highlighting various operational states of the system. The parameter $\xi$ captures the dependence on the magnetic field by representing a linear perturbation to both the conservative and non-conservative forces. The regions are classified according to the calculated response of the GMR modulation and are associated to the ratio between the radial oscillation frequency and the gyrotropic frequency. The details of these calculations are described in the "Effective model" subsection of the "Methods" section}.
  \label{fig:2}
\end{figure}

Magnetic textures include a diverse set of particle-like structures whose dynamics are described by equations \eqref{eq:EOMXY}. These structures include magnetic vortices, skyrmions, bubbles, and numerous other configurations \cite{gobel2021beyond}. In this case, $G_{i}$ is given by the emergent magnetic field and is associated to the topological winding number of the magnetic particle \cite{schulz2012emergent}. Magnetic vortices, for example, exhibit rigid particle behavior in the \unit[]{GHz} and lower frequency spectrum, associated with the lowest observed excitation modes within magnetic vortices \cite{guslienko2001field,guslienko2008magnetic,gaididei2010magnetic}. In the rigid particle behavior, we assume that the magnetization configuration and energy are uniquely defined by the position of the vortex core. To provide a physical realization of this system, we explore coupled vortices within a tri-layered nanopillar stack, shown in Fig. \ref{fig:2} (a). For the considered system, $G_{i} = -2\pi \Pi_{i}\Lambda_{i}$ and $D_{X\,i} = D_{Y\,i} = D_{\alpha\,i} = \alpha\pi \Lambda_{i}\ln\left(\mathcal{R}/l_{ex\, i}\right)$, where $i = 1,2$ refers to the bottom and top layers, respectively.\cite{guslienko2001field,guslienko2008magnetic,gaididei2010magnetic,dussaux2012field} $\Pi_{i}$ is the vortex polarity, $\Lambda_{i}$ is the layer thickness, $\mathcal{R}$ is the disk radius, $l_{ex\, i}$ is the exchange length and $\alpha$ is the  phenomenological Gilbert damping term. The self-potential $E_{0\,i}$ is associated with the constraining potential of the disk, which depends on the dipolar and exchange fields within each layer.
The dipolar interaction between the magnetic dipoles in the different layers generates the conservative interaction force between the two magnetic vortices. Due to the symmetries of the system, the conservative interactions must be a function of the relative distances between the two vortices, i.e. $r_{12} = \sqrt{\left(X_{1} - X_{2}\right)^2 + \left(Y_{1} - Y_{2}\right)^2}$. Near the equilibrium state, the interaction energy $E_{ij}$ takes on a harmonic potential form, expressed as $E_{int} = V_{12}r_{12}^2/2$, where $V_{12}$ depends on characteristic properties of the magnetic vortices, such as core size, chirality, and polarity. The dynamics of coupled magnetic vortices interacting via stray fields were previously studied in \cite{guslienko2005dynamics}, where only conservative couplings were analyzed. In contrast, this work also incorporates the effects of an applied electric current, which introduces two distinct phenomena. The magnetization in each layer locally polarizes the spin of the electrons. Once polarization occurs in one layer, it imparts a torque to the magnetization of the other magnetic layer through spin angular momentum transfer, known as spin transfer torque (STT) \cite{gaididei2010magnetic}. The polarization of the electric current in each layer and the STT depend on the magnetic configuration of each layer. Additionally, the current generates an Oersted field that depends on its direction. The Oersted field can change the frequency of the magnetic vortex. For a vortex where the external parameters are fixed, the magnetization in each layer is uniquely defined by its core position alone. As a result, the scattering of electrons creates a non-conservative interaction between the two vortices that, in the rigid particle assumption, depends on their relative core positions. External parameters such as magnetic or electric fields, spin-polarized currents, and temperature variations can modify the magnetic vortex profiles. Using these external perturbations as control parameters allows for the modulation of forces and enlarges the phase space of the system, providing a richer set of functional responses. In this work, we consider a modulation of the VC size by an applied magnetic field, $\mu_0 H_{\perp}$, applied perpendicularly out-of-plane with respect to the layers. For details, refer to the "Effective model" subsection of the "Methods" section.

We study the system presented in Fig. \ref{fig:2} (a). Both the bottom and top layers (layers 1 and 2) are modeled as Permalloy (Ni$_{81}$Fe$_{19}$) with thicknesses \unit[15]{nm} and \unit[4]{nm}, respectively, with a copper spacer in between them (thickness of \unit[10]{nm}). As such, the considered system is a Giant Magneto-Resistance (GMR) spin-valve. At remanence, the two magnetic layers exhibit vortex configurations referred to as V$_{1}$ and V$_{2}$. The vortex chiralities are set to be clockwise, aligning with the direction of the Oersted field generated as the current flows from the top layer (layer 2) to the bottom one (layer 1). Vortex polarities, $\Pi _{1} ^-$ and $\Pi _{2} ^+$, are chosen to be anti-parallel, thus adopting the case where VC auto-oscillations lead to conveniently detectable high-frequency voltage generation \cite{locatelli2011dynamics,hamadeh2014origin}.

To generate the non-conservative coupling $F_{ij}$ and drive the system out of equilibrium, we consider a direct current $I_{\mathrm{dc}}$ applied to the structure which is exerting spin transfer torques on both vortices. Once $I_{\mathrm{dc}}$ exceeds a certain threshold, it induces auto-oscillations of the gyrotropic mode of the magnetic VC. This process allows us to fine-tune the operating frequencies and coupling strength by adjusting the applied dc current.  Additionally, a control parameter, $\xi$, is introduced to assess the strength of the coupling strengths, both conservative and non-conservative. This coupling parameter has an analogous effect to an applied magnetic field that influences the size of the VC. For more details, see the "Effective model" subsection of the "Methods" section.

\subsection*{Classification of the orbits}
The numerical calculations based on Eq.~(\ref{eq:EOMXY}) consistently reveal distinctive behaviors for the vortices in layers 1 and 2. Both vortices exhibit a periodic angular motion with a frequency $\omega_{g}$. However, notable differences emerge in their radial characteristics: the vortex core (VC) in layer 2 is confined to smaller radii, whereas the VC in layer 1 follows trajectories with larger radii. Additionally, the orbit of the layer 2 VC displays radial oscillations at a frequency $\omega_{r}$. While similar radial oscillations might be anticipated for the VC in layer 1, these are significantly less pronounced. This disparity arises due to the non-conservative coupling between the two vortices.

Fig.~\ref{fig:2}(b) illustrates various orbits corresponding to different values of $I_{\mathrm{dc}}$ and the control parameter $\xi$, which accounts for the influence of the magnetic field. In all cases shown, the vortex core (VC) in layer 1 follows a circular, well-centered trajectory with a gyrotropic frequency $\omega_{g}$. In contrast, the orbits of the VC in layer 2 exhibit a combination of angular motion at frequency $\omega_{g}$ and radial oscillations at frequency $\omega_{r}$. By tuning the interactions between the vortices, specifically through $I_{\mathrm{dc}}$ and $\xi$, it is possible to adjust the frequencies such that the ratio $\omega_{r} / \omega_{g}$ assumes a rational value. We define this rational value as $q = \omega_{r} / \omega_{g}$. Due to the difference in operational frequencies, we introduce the concept of states to uniquely classify each case. For instance, the state $\Psi_{0}$ corresponds to $q = 0$ where the VC in layer 2 undergoes a circular trajectory that shares the same center with that of the VC in layer 1. $\Psi_{1}$ corresponds to $q = 1$ where the VC in layer 2 undergoes a circular trajectory, but with slightly shifted center. The state $\Psi_{2}$ corresponds to $q = 3/2$. In this case,  the VC of layer 2 undergoes a rather complex shaped trajectory. The superposition of $q = 1$ and $q = 2$ is regarded as state $\Psi_{3}$, and $q = 2$ is denoted by $\Psi_{4}$. In the state $\Psi_{4}$, layer 2's VC undergoes a somewhat non-circular trajectory, whereas it undergoes a shifted one in state $\Psi_{3}$.  Fig. \ref{fig:2} (c) presents a phase diagram as an illustration of the frequency of radial oscillation of layer 2's VC with respect to $I_{\mathrm{dc}}$ and $\xi$. The regions of the phase space are characterized by the aforementioned radial oscillation frequency corresponding to fractional harmonics of the gyrotropic mode's frequency. The boundary of these regions is intricately defined by both the current and the coupling parameter $\xi$, revealing a complex behavior.

\subsection*{Nonlinearity of the coupled dynamics}
In the presented model addressing the coupled vortices, we have considered a minimal approach to describe the conservative and non-conservative forces, taking into account an expansion up to first order in the distance between the vortices. We noticed that the potential energy for the magnetic texture due to exchange coupling and stray fields within the disk is inherently nonlinear. We found that introducing a non-conservative force drives the system into the nonlinear regime. This occurs because the frequency is contingent on the radial distance of the VC, and the coupling introduces variations in these radial distances. In contrast to conventional methods in the literature, which typically necessitate higher-order perturbations of the damping factor \cite{dussaux2012field}, our model directly derives this nonlinearity from the interplay between the two vortices up to linear order. It is clear that by considering higher order perturbations in both the forces and damping factors, a more pronounced nonlinear behavior will be observed. For additional details and supporting data, including a description of the Effective model, refer to the "Effective model" subsection of the "Methods" section.
Since the system is driven in the nonlinear regime, it becomes highly responsive to variations in parameters. As a result, the region boundaries and expected orbits shown in Fig. \ref{fig:2} (c) can be changed significantly with parameter changes.

\subsection{Micromagnetic Simulations}
\begin{figure*}[!ht]
  \centering
  \includegraphics[width=0.6\textwidth]{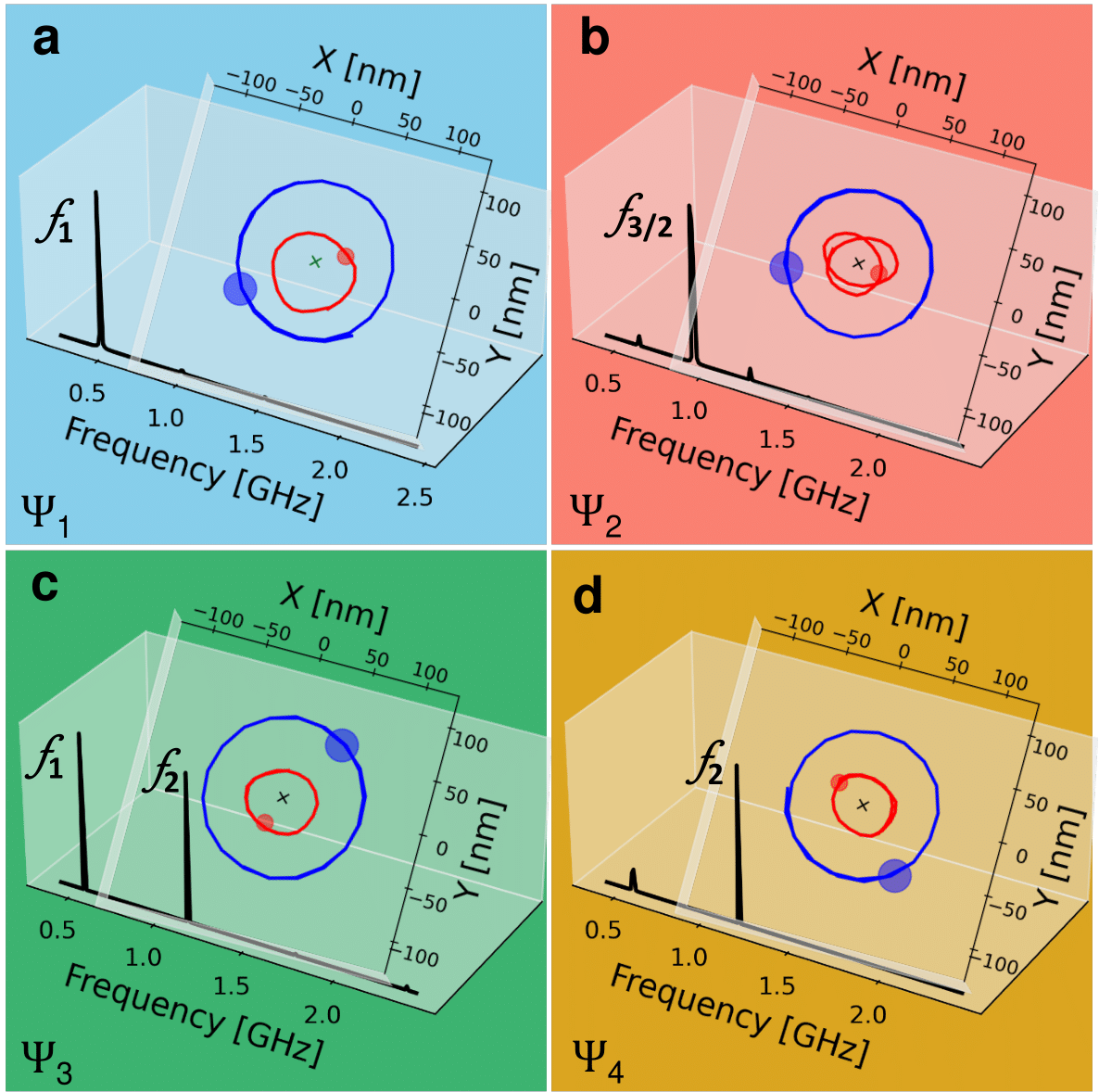}
 \caption{\textbf{Micromagnetic simulations}: (a, b, c, d) Trajectories undergone by the vortex cores (VCs) are traced with a blue line and dot for the VC in layer 1 and with a red line and dot for the VC in layer 2. Fast Fourier Transform (FFT) of the GMR spectra at fixed $H_{\perp}$ and $I_{\mathrm{dc}}$, corresponding to the four highlighted states (the colored background matches the colored labels of Fig. \ref{fig:2}), are also plotted for each state. The states correspond to the following parameter values: (a) $\Psi_{1}$ state at $I_{\mathrm{dc}} = 10 , \mathrm{mA}$ and $\mu_0 H_{\perp} = 1 , \mathrm{mT}$, (b) $\Psi_{2}$ state at $I_{\mathrm{dc}} = 20 , \mathrm{mA}$ and $\mu_0H_{\perp} = 10 , \mathrm{mT}$, (c) $\Psi_{3}$ state at $I_{\mathrm{dc}} = 20 , \mathrm{mA}$ and $\mu_0H_{\perp} = 50 , \mathrm{mT}$, and (d) $\Psi_{4}$ state at $I_{\mathrm{dc}} = 20 , \mathrm{mA}$ and $\mu_0 H_{\perp} = 50 , \mathrm{mT}$.}
  \label{fig:3}
\end{figure*}

To validate the analytical results, we perform full micromagnetic simulations of the coupled vortices system presented earlier by means of the software package FastMag \cite{chang2011fastmag}. FastMag is a finite element micromagnetic simulator that solves the Landau-Lifshitz Gilbert (LLG) equation at every cell of the discretised system. For further details regarding the micromagnetic simulations, refer to the "Micromagnetic simulations" subsection of the "Methods" section.

In Fig. \ref{fig:3} (a, b, c, d), we present the trajectories underwent by both layers' VCs for fixed perpendicular magnetic field $\mu_0H_{\perp}$ and dc current $I_{\mathrm{dc}}$ values. $I_{\mathrm{dc}}$ excites the gyrotropic mode and tunes the non-conservative coupling via the generated Oersted field, whilst $\mu_0H_{\perp}$ tunes the conservative coupling and alters the fundamental gyration frequencies. When compared to the analytical results, these trajectories can be identified as the previously introduced states $\Psi_{1}$, $\Psi_{2}$, $\Psi_{3}$, and $\Psi_{4}$. Hence, the results from our simulations confirm that the analytical model of Eqn.~\ref{eq:EOMXY} is describing the vortex system. Additionally, the Fast Fourier Transform (FFT) spectra of the calculated GMR signal highlight the gyrotropic mode's frequency of the respective states at fixed $\mu_0H_{\perp}$ and $I_{\mathrm{dc}}$. It can be inferred from the GMR power spectra that different harmonics of the gyrotropic mode are excited, whereby their frequencies and power depend on current and field. In fact, the strength of the different harmonics enables to distinguish between the different states $\Psi$ of the system since the relative distribution of the power in $f_1$, $f_2$ and $f_{3/2}$ is specific for every state. These different harmonics are defined as the base frequency $f_1$, which can be identified as the fundamental gyration frequency of the VC in the thick layer, its second harmonic $f_2$ which refers to double the fundamental frequency, and its half-integer harmonic $f_{3/2}$ which refers to three halves the fundamental frequency. 

\begin{figure*}[!ht]
  \includegraphics[width=\columnwidth]{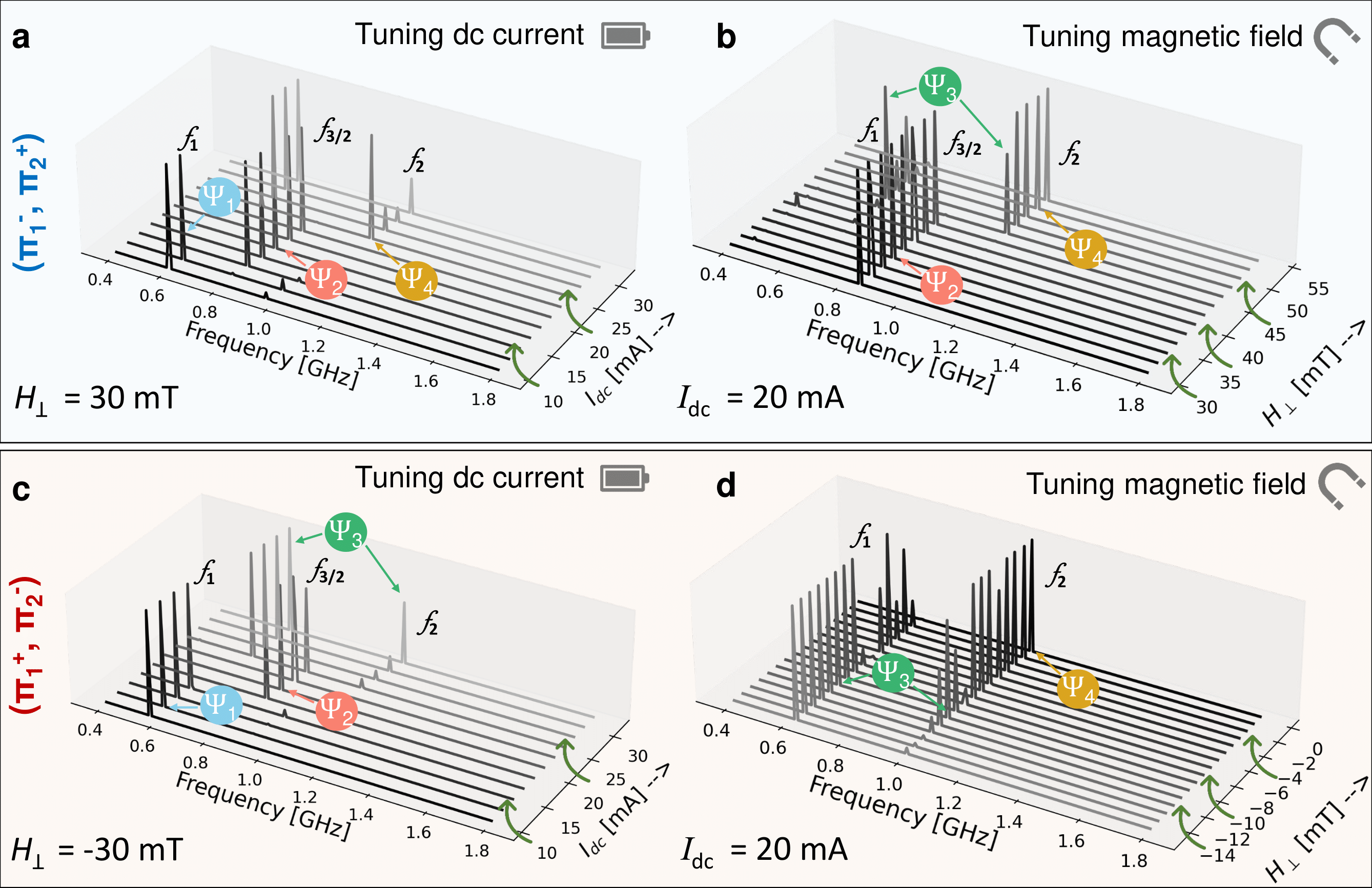}
 \caption{\textbf{Micromagnetic Simulations:} (a, b, c, d) FFT Power spectra extracted for current and/or field tuning presented for the two configurations of vortices with opposite polarities ($\Pi _{1} ^-$, $\Pi _{2} ^+$ (a, b) and $\Pi _{1} ^+$, $\Pi _{2} ^-$ (c, d)). Depending on the field /current combination, different VC orbit states $\Psi_{i}$ are realized which are characterized by the relative weights of the oscillations at $f_1,f_2$ and $f_{3/2}$.}
  \label{fig:4}
\end{figure*}

  To illustrate how the different states can be realized, we perform current/field sweeps, whereby one of which is swept while the other is kept constant (see Fig. \ref{fig:4} (a, b, c, d)). The notations $\Pi _{i} ^j$ presented on the left side of Fig. \ref{fig:4} refer to the vortex polarity in either magnetic layer, where $i$=1, 2 represents the layer itself, and $j$=$+$, $-$ refers to the polarity of the vortex in that layer. In Fig. \ref{fig:4} (a, b), we present the first case of opposite vortex polarities with the bottom layer having a negative polarity and the top layer having a positive one, namely ($\Pi _{1} ^-$, $\Pi _{2} ^+$). We first perform a current sweep at a fixed magnetic field $\mu_0H_{\perp}$ = \unit[30]{mT} applied perpendicular to the structure's plane (see Fig. \ref{fig:4} (a)). The FFT spectrum consists of different harmonics of the gyrotropic mode's fundamental frequency $f_1$. These harmonics are marked as $f_{3/2}$ and $f_{2}$. Increasing $I_{\mathrm{dc}}$ from \unit[10]{mA} to \unit[15]{mA}, the signal starts with most power at the fundamental frequency $f_1$, and changes to $f_{3/2}$. Increasing $I_{\mathrm{dc}}$ further, the signal has most power at $f_2$. Beyond \unit[25]{mA}, the FFT signal goes back to showing most power for $f_1$. As the analysis of the VC trajectories shows, this change in the power dominance of the various harmonics is due to a change in the system's state. For low current, the system started in state $\Psi_{1}$, transitioned to $\Psi_{2}$, to $\Psi_{4}$ and back to $\Psi_{1}$ for the swept current range. In Fig. \ref{fig:4} (b) we sweep $\mu_0H_{\perp}$ while keeping $I_{\mathrm{dc}}$ fixed at \unit[20]{mA}. For field values below \unit[45]{mT}, the FFT signal has most power at $f_{3/2}$, which means that the system is in state $\Psi_{2}$. Increasing $\mu_0H_{\perp}$, the signal attains maximum power mostly at $f_{2}$ meaning a transition from $\Psi_{2}$ to $\Psi_{4}$, except for $\mu_0H_{\perp}$ = \unit[45]{mT} where the FFT has about equal power for $f_1$ and $f_2$, which means that the system is in state $\Psi_{3}$. Choosing the alternate configuration ($\Pi _{1} ^+$, $\Pi _{2} ^-$) (see Fig. \ref{fig:4} (c, d)), the same approach is followed where we first swept $I_{\mathrm{dc}}$ and held $\mu_0H_{\perp}$ fixed at \unit[-30]{mT}, then fixed $I_{\mathrm{dc}}$ at \unit[20]{mA} and swept $\mu_0H_{\perp}$. Similar results are observed whereby a change in the FFT signal's power between the different harmonics is achieved for different values of the control parameters..

\subsection{Experimental Results}
\begin{figure*}[!ht]
  \includegraphics[width=\columnwidth]{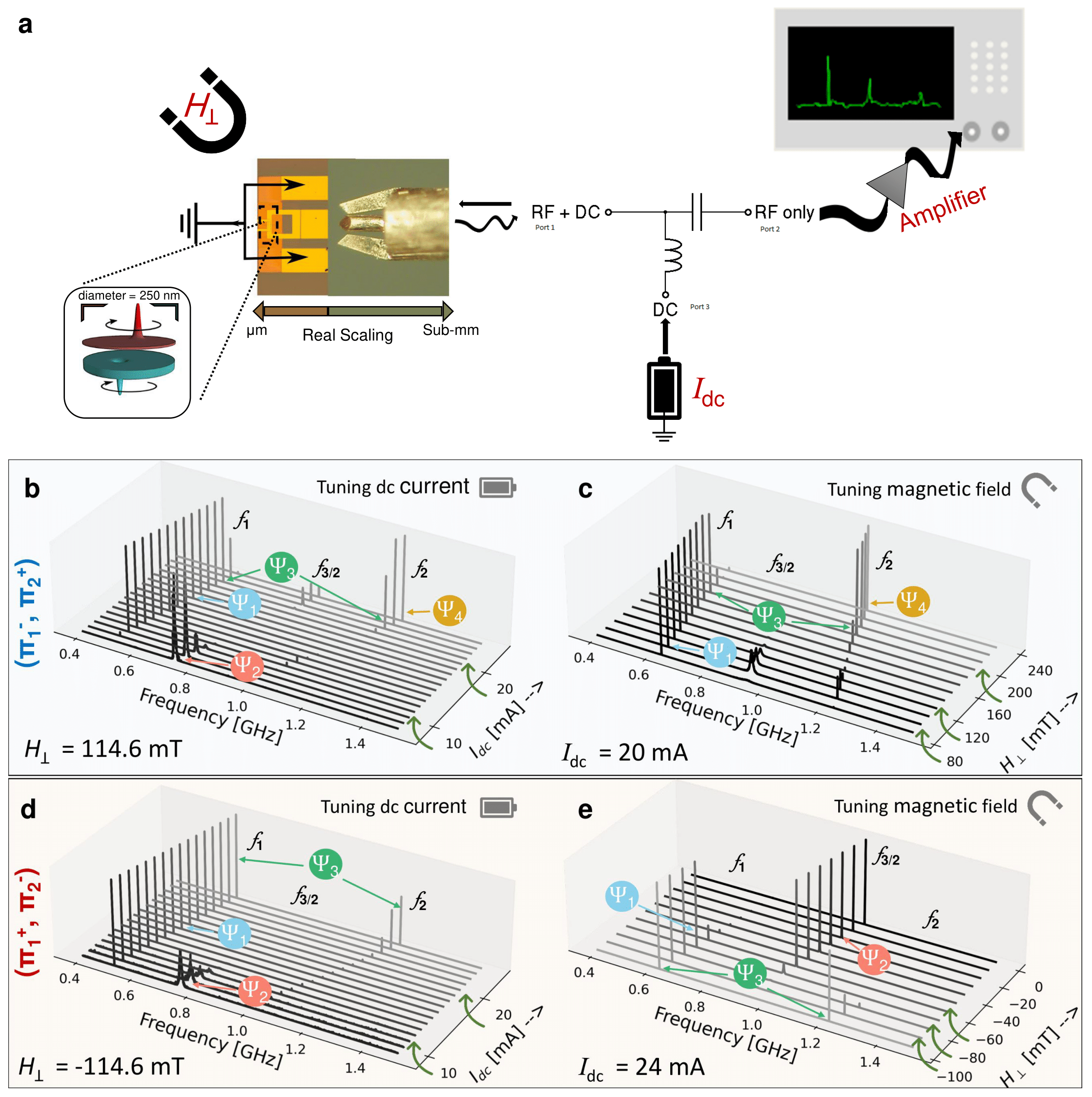}
 \caption{\textbf{Experimental results:} (a) Schematic of the experimental setup in which the dynamics of the coupled vortices are monitored electrically through their GMR emission spectra. To realize this, a Bias-T is used to inject dc current into the structure, and to direct back the generated rf output through an amplifier and onto a spectrum analyzer for subsequent visualization. The picoprobe, contacts and structure are overscaled for presentation purposes. (b ,c) Normalized power spectra of the measured GMR signal for a vortex with negative polarity in layer 1 ($\Pi _{1} ^-$) and a vortex with positive polarity in layer 2 ($\Pi _{2} ^+$). (d, e) Power spectra of the measured GMR signal for a vortex with positive polarity in layer 1 ($\Pi _{1} ^+$) and a vortex with negative polarity in layer 2 ($\Pi _{2} ^-$). $\mu_0H_{\perp}$ is fixed and$I_{\mathrm{dc}}$ is tuned (b,d) and vice versa (c, e).}
  \label{fig:5}
\end{figure*}

To demonstrate the realization of the different VC orbit states, we measure a nanopillar which hosts the proposed coupled vortices system. Both the bottom and the top layers are made of Permalloy (Ni$_{81}$Fe$_{19}$) with thicknesses \unit[15]{nm} and \unit[4]{nm}, respectively, whereas the copper spacer in between has a thickness of \unit[10]{nm}. As such, the considered system is a Giant Magneto-Resistance (GMR) spin-valve (further details on the sample can be found in the "Sample" subsection of the "Methods" section). At remanence, vortex configurations are selected in agreement with those from the analytical model. As such, the vortex ground states have clockwise chiralities and anti-parallel polarities. Furthermore, the current is also chosen as directed from the top layer to the bottom one.

In Fig. \ref{fig:5} (a), we present the experimental setup to investigate the GMR signal. A dc current is injected through a Bias-T across the structure. We tune the operating frequencies and the coupling strength of the magnetic vortices by varying the applied dc current, $I_{\mathrm{dc}}$ and/or the perpendicular magnetic field, $\mu_0H_{\perp}$. $\mu_0H_{\perp}$ plays a crucial role in changing both the size of the VC, and hence the coupling between both VCs. Additionally, $\mu_0H_{\perp}$ alters the characteristic gyrotropic frequency of the VC. We monitor the amplified GMR signal using a spectrum analyzer (Additional information can be found in the "Microwave measurement setup" subsection of the "Methods" section). The probed GMR signal for the two possible configurations with anti-parallel vortex polarities is presented in Fig. \ref{fig:5} (b, c, d, e). In the first configuration ($\Pi _{1} ^-$, $\Pi _{2} ^+$) (see Fig. \ref{fig:5} (b, c)), the vortex of the bottom layer has a negative polarity, and that of the top layer has a positive one. To illustrate the experimentally realized states, in Fig. \ref{fig:5}(b), we fix $\mu_0H_{\perp}$ and tune $I_{\mathrm{dc}}$. In this case, $f_{3/2}$ dominates for low currents, $f_{1}$ for intermediate currents whereas both $f_{1}$ and $f_{2}$ are pronounced for high currents, except at $I_{\mathrm{dc}}$ = \unit[25]{mA} where $f_{2}$ dominates. This can be understood as a transition between different states for a fixed field and tunable current. According to the observations illustrated in Fig. \ref{fig:3}, the system transitions between the states $\Psi_{2}$, $\Psi_{1}$, $\Psi_{3}$ and  $\Psi_{4}$ respectively. To demonstrate the fact that the coupled dynamics can be tuned by both $I_{\mathrm{dc}}$ and $\mu_0H_{\perp}$, we fix $I_{\mathrm{dc}}$  and tune $\mu_0H_{\perp}$ (Fig. \ref{fig:5}(c)). For the measured field values, $\Psi_{1}$ dominates for low to intermediate field values  between $\mu_0H_{\perp}$ = \unit[80]{mT} and \unit[180]{mT}, then a transition to $\Psi_{3}$ takes place between $\mu_0H_{\perp}$ = \unit[180]{mT} and \unit[210]{mT}. For fields above \unit[210]{mT} a transition to $\Psi_{4}$ is observed.

Studying the alternate configuration ($\Pi _{1} ^+$, $\Pi _{2} ^-$) (Fig. \ref{fig:5} (d, e)), we first fix $\mu_0H_{\perp}$ at \unit[-114.6]{mT} and tune $I_{\mathrm{dc}}$ from \unit[8]{mA} to \unit[26]{mA}. As expected, this leads to varying the frequency and strength of the GMR harmonics in a manner qualitatively analogous to that shown in the first configuration.


\begin{figure*}[!ht]
  \includegraphics[width=\columnwidth]{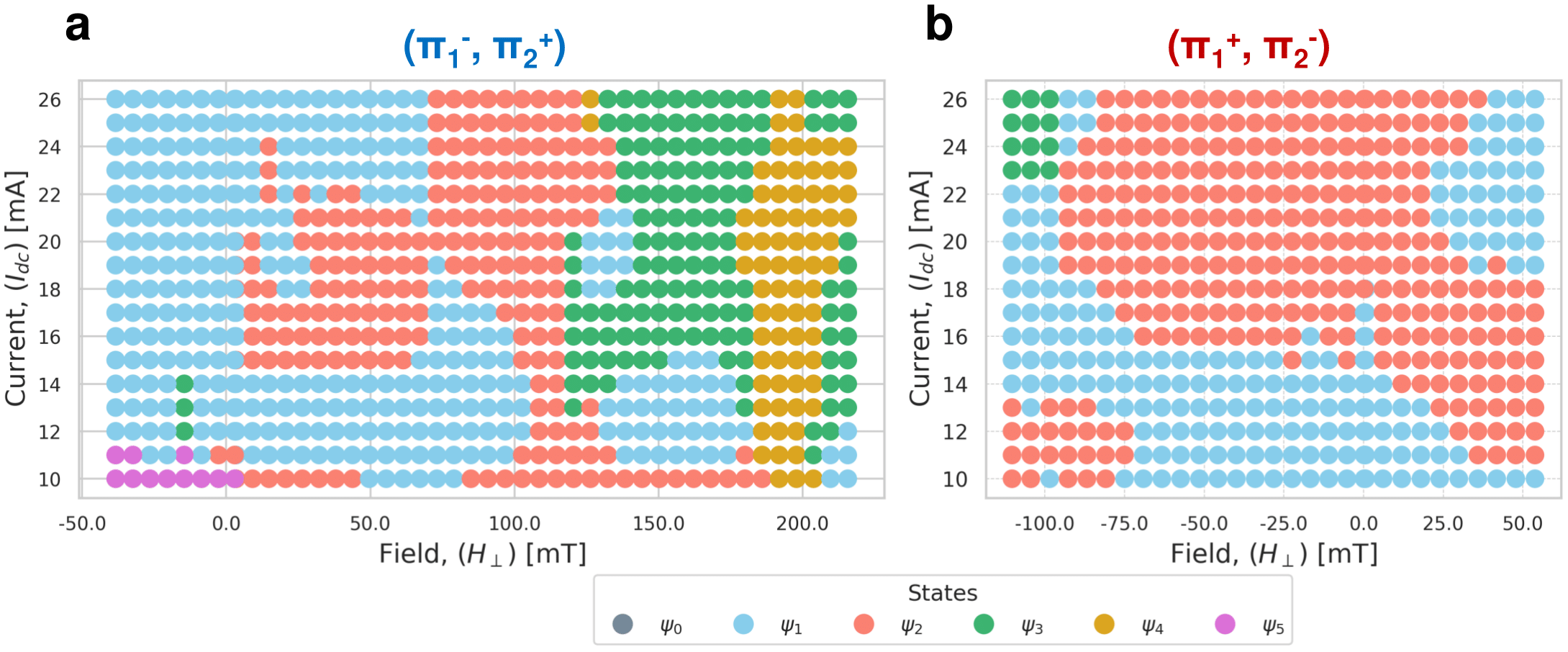}
 \caption{\textbf{Experimental phase space:} System state representation as a function of external field $\mu_0 H_\perp$ [mT] and current $I_\mathrm{DC}$ [mA] for the two configurations (a) ($\Pi _{1} ^-$, $\Pi _{2} ^+$) and (b) ($\Pi _{1} ^+$, $\Pi _{2} ^-$). States have been identified via their characteristic harmonics in the GMR spectra (compare to Fig.~\ref{fig:4} and Fig.~\ref{fig:5}).}
  \label{fig:6}
\end{figure*}

From these measurements, we can conclude that we have successfully tuned the conservative and non-conservative coupling in our system with the external magnetic field and the applied current. This leads to qualitatively different VC trajectories manifesting in qualitatively different GMR spectra. This allows for a systematic classification of the experimentally achieved VC orbit states in a phase map. As such, we associate configurations ($\Pi _{1} ^-$, $\Pi _{2} ^+$) and ($\Pi _{1} ^+$, $\Pi _{2} ^-$) with maps that serve to determine the previous, present, and future state of the system just by knowing the input $I_{\mathrm{dc}}$ and $\mu_0H_{\perp}$ values fed to the system (see Fig. \ref{fig:6} (a, b)). In Fig. \ref{fig:6} (a), we present such a phase map for a range of $I_{\mathrm{dc}}$ and $\mu_0H_{\perp}$ values. The possibility of transitioning between states enhances the richness of the system, as each exhibits a different operational frequency. Varying the applied current strengthens/weakens the non-conservative coupling between the two gyrating VCs. The current induced Oersted field and the external magnetic field yield change in the fundamental frequency of the system. In Fig. \ref{fig:6} (b), we present the alternate configuration of opposite VCs' polarizations. Here, increasing $\mu_0H_{\perp}$ values are mostly applied along the negative $z$-direction, which coincides with the polarization of the top layer's VC. Once again, the state is determined by the input current and/or field, where the state manipulation can be achieved by varying the input.


\section{Discussion}\label{sec5}

\begin{figure}[!ht]
  \includegraphics[width=\columnwidth]{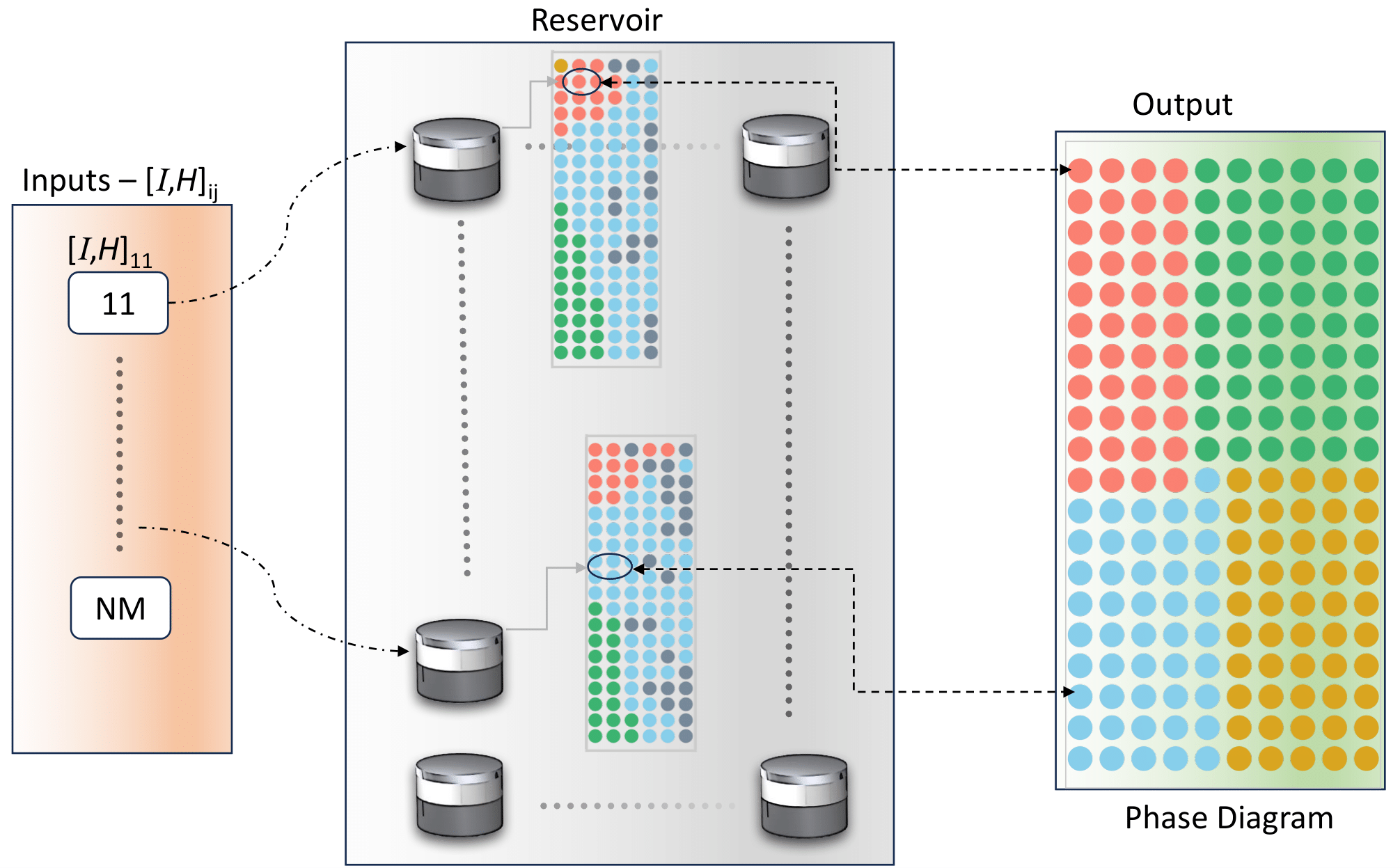}
 \caption{Schematic of a reservoir computer harnessing the capabilities of the system under investigation. In this envisioned configuration, the reservoir comprises a series of devices, each of which can be composed of $N_{\mathrm{i}}$ layers of interconnected vortices. The applied current to each device functions as an input signal. Consequently, the reservoir generates a dynamic phase diagram akin to the one depicted in Fig. \ref{fig:6}. The characteristics of this phase diagram are shaped by the specific values of the applied current $I_{\mathrm{i}}$ and magnetic field $\mu_0H_{\mathrm{j}}$ for each individual device. The two exemplary phase diagrams, presented for two devices, are illustrative examples and were computed using the effective model for two trilayers with the following structures: 12(Py)/8(Cu)/4(Py) and 10(Py)/10(Cu)/5(Py), where the numbers represent the layer thickness}
  \label{fig:7}
\end{figure}

A characteristic of a reservoir-based computer is its ability to classify using a nonlinear function of its inputs \cite{maass2002real,tanaka2019recent}. In the reservoir computing paradigm, input information propagates through a highly nonlinear dynamical system with short-term memory called the reservoir. Due to its own complex dynamics, the reservoir projects the input data into a high-dimensional space represented by the output layer. Classification is then performed on this high-dimensional representation using simple linear regression. Fig. \ref{fig:7} shows a schematic of how the presented vortex auto-oscillator can be used in the context of reservoir computing. The setup involves a collection of coupled vortices auto-oscillators, each of which is capable of projecting an input current into the complex electrically readable phase space, similar to Fig. \ref{fig:6}. By combining the outputs of these devices, simple linear regression allows for efficient classification tasks. The output of each device is based on the relative orbits of the individual vortices, which are influenced by their mutual interactions associated to their relative distances.  Despite the promising potential of coupled vortex systems in reservoir computing, several challenges must be addressed for practical implementation. One key challenge is the precise control and reproducibility of vortex dynamics. The sensitivity of the system to small variations in external fields and currents introduces complexity in maintaining stable operational states, which can limit scalability in real-world applications. Additionally, the inherent nonlinearity and chaotic behavior of the system, while advantageous for complex computations, presents difficulties in ensuring consistent and predictable performance across multiple devices. As a result, an element of memory is naturally exhibited within the vortex configuration of each stack, with a non-trivial dependence on both input currents and vortex configurations. The proposed device offers higher non-linearity and tunability than current spin-torque nano-oscillators schemes due to the added degrees of freedom \cite{torrejon2017neuromorphic,markovic2019reservoir,kanao2019reservoir} associated with the complex trajectories induced by the coupling between the VCs. Adding to the increased non-linearity and tunability, higher output powers are achievable by substituting the utilized GMR spin-valves with tunnel magnetoresistance (TMR) junctions. The TMR-based devices provide significantly higher output powers, making them more suitable for practical applications and potential implementation in advanced systems.  Fabrication-related challenges also pose significant hurdles. The variability in material properties, such as Gilbert damping coefficients and exchange stiffness, can lead to deviations in the expected dynamical behavior of the system, requiring further optimization and standardization of deposition techniques. Furthermore, the interaction of vortices in multilayered structures is influenced by factors such as layer thickness variations and interface effects, which can impact coupling strength and overall performance. Addressing these challenges necessitates a combination of refined fabrication processes and advanced characterization techniques. Perhaps another major step to further elevate the system's non-linearity and diversity is to utilize a $N$-layered stack instead of the tri-layered one. In this concept, the complexity and diversity of the output depends on the properties of each layer, and the ability to tune the magnetic field. By introducing this tunability, this approach enables a highly versatile reservoir computer capable of performing classifications on different sets simply by adjusting the field appropriately. Another critical issue is the energy efficiency and power dissipation of the system. While magnetic vortex oscillators offer potential advantages in terms of low power consumption, the requirement for precise control of non-conservative coupling through applied currents can introduce additional power overhead. Strategies such as optimizing the design of the nanopillar stack and leveraging advanced materials with lower damping properties could help to mitigate these concerns. This is especially interesting if mechanisms like voltage-controlled magnetic anisotropy \cite{amiri2012voltage} or micromagnets mounted on micro- or nano mechanical systems \cite{cocconcelli2024tuning} are used which allow to control the field locally, so individually for every oscillator. This tunability of the reservoir allows control over its dynamic response, allowing for different projections into the output layer.  Overall, while the proposed system presents a highly tunable and nonlinear computational platform, achieving practical deployment will require addressing the aforementioned challenges through interdisciplinary efforts in materials science, nanofabrication, and computational modeling. Future research efforts should focus on developing scalable manufacturing processes, improving device robustness, and exploring hybrid integration with other emerging computing paradigms, such as neuromorphic and quantum computing. As a result, the proposed reservoir can perform multiple tasks, overcoming a major limitation of traditional reservoir computing, which typically is restricted to a single task.

\section{Conclusion}\label{sec6}

In conclusion, starting from a general analytical model for the coupled motion of VCs, we have predicted and experimentally demonstrated that the conservative and non-conservative coupling contributions lead to complex VC orbits. These orbit states depend strongly non-linearly on the external control parameters given by the external magnetic field and the applied current. Since the individual orbit states have a characteristic fingerprint in their GMR power spectra showing several harmonics of the basic VC gyration frequency, they can be easily read-out electronically from the investigated spin-valve nanopillar. Given the high non-linearity and tunability of the system, we discuss the perspective role of the nano vortex oscillator as a building block for a flexible reservoir computer and how the non-linearity of the system can be further increased and tailored by the use of more than two coupled magnetic vortices.

\section{Methods}\label{sec7}

\subsection{Sample}

The spin-valve nanopillar used in this study consists of a permalloy (Ni$_{80}$Fe$_{20}$) bilayer separated by a \unit[10]{nm} copper (Cu) spacer, deposited on a silicon (Si) wafer. The permalloy layers are \unit[4]{nm} and \unit[15]{nm} in thickness. The nanopillar was fabricated with a diameter of \unit[250]{nm} using established e-beam lithography and ion milling techniques

\subsection{Microwave measurement setup}
For the microwave electrical measurements detailed in this study, the procedures were conducted at room temperature, employing the circuit configuration illustrated in Fig. \ref{fig:5} (a) of our manuscript. The specific setup involved mounting the sample in a stationary holder placed between the poles of an electromagnet. In our experimental arrangement, a steady direct current was directed through the device under test. Concurrently, the alternating current component was isolated using a bias-T setup. This extracted signal was then subjected to amplification for clarity. The analysis of this signal was carried out using a spectrum analyzer. To enhance the signal quality, a low noise preamplifier was also incorporated into the system. This preamplifier features a 32 dB gain, a low noise figure of 1.9 dB, and operates effectively in the \unit[4]{}-\unit[20]{GHz} frequency range. In the framework of our experimental measurements,  $\mu_0H_{\perp}$ was applied in the positive $z$-direction for ($\Pi _{1} ^-$, $\Pi _{2} ^+$), which is the direction of the top layer's VC polarity (see Fig. \ref{fig:2} ($\Pi _{1} ^-$, $\Pi _{2} ^+$)). For ($\Pi _{1} ^+$, $\Pi _{2} ^-$),  $H_{\perp}$ was applied in the negative $z$-direction (see Fig. \ref{fig:5} ($\Pi _{1} ^+$, $\Pi _{2} ^-$)), which is the direction of the top layer's VC polarity in this case. This is important to consider because of the layer asymmetry, where the thin layer's VC is more susceptible to switching has a magnetic field of sufficient amplitude opposed its polarity. Furthermore, in reference to Fig. \ref{fig:6} (b), the appearance of only three states in the respective map of configuration ($\Pi _{1} ^+$, $\Pi _{2} ^-$) is expected due to the limited experimental data. Increasing $\mu_0H_{\perp}$ beyond \unit[-110]{mT}, higher indexed states such as $\Psi_{4}$ are expected to appear.

\subsection{Micromagnetic simulations}
In regards to the micromagnetic simulations performed via FastMag, FastMag uses the finite element method (FEM) to solve the Landau Lifshitz Gilbert Slonczewski (LLGS) equation (see the "Effective model" subsection of the "Methods" section). FEM is particularly useful for simulating complex shaped systems with a non-regular grid, such as those considered here. The magnetic parameters used in our simulations are as follows, the saturation magnetization of the bottom Permalloy layer is \unit[7.6$\times$10$^5$]{A $\mathrm{m^{-1}}$}, and that of the top layer is \unit[6.4$\times$10$^5$]{A/m}. The Gilbert damping coefficient of the bottom Permalloy layer is \unit[0.9$\times$10$^{-2}$]{}, and that of the top layer is \unit[1.4$\times$10$^{-2}$]{}. These parameters were taken from the experimental work \cite{naletov2011identification}. The exchange stiffness for our Permalloy ferromagnets was obtained from literature, and is $A$ = \unit[1.0$\times$10$^{-11}$]{J $\mathrm{m^{-1}}$}. The magnetization $M\text{(x,y,t)}$ and resistance were computed over a period of \unit[1]{$\mu$s}. The microwave frequency was calculated by performing a Fourier transform from \unit[500]{ns} to \unit[1]{$\mu$s}, which is long enough to resemble the condition of dynamic steady state. To excite the auto-oscillation dynamics, $I_{\mathrm{dc}}$ is applied downwards through the structure. The Oersted field generated by the current is pre-computed and added as an applied field component. A positive current corresponds to a flow of electrons from the bottom layer to the top one. Upon the injection of an above threshold $I_{\mathrm{dc}}$ through the structure, the gyrotropic mode is excited by STT \cite{pauselli2015spin}.

\subsection{Effective  model}

The magnetization dynamics are well described by the  LLGS  equations \cite{guslienko2008magnetic}:
\begin{align}
\frac{d\vec{m}_i}{dt} &=-\gamma_0 \,  \vec{m}_i \times \vec{H}_i + \alpha_i \,  \vec{m}_i \times \frac{d\vec{m}_i}{dt} - \sigma_i \, \vec{m}_i\times \left( \vec{m}_i \times \vec{m}_j \right)
\end{align}
where $i=1,2$ refers to the bottom and top layers, respectively, and $j\neq i$. We consider the spin-polarized current in the bottom layer being produced as a back scattering from the top layer, thus $\sigma_i= \pm g \mu_B J P_i/2 e M_{s\, i} \Lambda_i$, where the top/lower sign holds for the top/bottom layers, respectively. $\gamma_0,\mu_B,e,g,J$ are the Gilbert gyromagnetic ratio, Bohr magneton, electron charge, Landé factor (positives values) and current density, respectively, while $\Lambda_i, M_{s\, i}, P_i$ are the magnetic layer thicknesses, saturated magnetization on each layer, and the polarization of the spin-currents to interact with each layer. Furthermore, magnetic interactions within each layer due to stray fields, and the Oersted field due to charge current are included in the effective field $\vec{H}_i$.\\
The coordinates of magnetic vortex cores behave as collective coordinates and obey the Poisson bracket relation $\{X, Y\} \propto 1$.\cite{guslienko2001field,guslienko2008magnetic,gaididei2010magnetic} Consequently, the dynamics of the vortex core positions can be described by Eq.~\eqref{eq:EOMXY}. To derive these equations in the context of coupled vortices in multilayers, we utilize the Thiele formalism~\cite{thiele1973steady}. This is achieved by projecting the LLGS equations onto $\left(\vec{m}_i \times \partial_a \vec{m}_i\right)$ for each layer, where $\partial_a$ represents spatial derivatives. Performing a spatial integration for each layer yields the following equations
\cite{guslienko2001field,guslienko2008magnetic,gaididei2010magnetic} \begin{subequations}\label{eq:EOMpolar} \begin{equation} G_{i}s_{i}\dot{\phi_{i}} = -D_{\alpha\,i}\dot{s_{i}} - \frac{\partial E_{0}}{\partial{s_{i}}} -  \frac{\partial E_{int}}{\partial{s_{i}}} + F_{s_{i}} \end{equation} \begin{equation} G_{i}\dot{s}_{i} = D_{\alpha\,i}s_{i}\dot{\phi_{i}} + \frac{1}{s_{i}}\frac{\partial E_{0}}{\partial{\phi_{i}}} + \frac{1}{s_{i}}\frac{\partial E_{int}}{\partial{\phi_{i}}} - F_{\phi_{i}}, \end{equation}
\end{subequations}
where, here we considered the position of the vortices' core in polar coordinates $s$, $\phi$ to show explicitly the cylindrical symmetry. In the $X,Y$ coordinates, we obtain back the equations in \eqref{eq:EOMXY}.
The magnetic energy $E_{0}$ includes the confinement potential resulting from magnetic interactions within the layer, in particular the dipolar and exchange interactions, as well as the influence of the magnetic field induced by the electric current (Oersted field). These interactions are well documented in the literature.\cite{guslienko2001field,guslienko2008magnetic,gaididei2010magnetic} Conversely, the interacting energy $E_{int}$ is related to the dipolar coupling between magnetic dipoles in the two layers. In the realm of non-collinear magnetic textures, such as magnetic vortices, this coupling can present considerable complexity. However, an effective theoretical approach allows us to streamline this complexity by considering the dependence of the potential only on the distance between the two vortices, thus simplifying it to \cite{rodrigues2017spin}:
\begin{align}
E_{\text{int}} &=  V_{ji}(|\vec{s}_i-\vec{s}_j|)\nonumber\\
&\approx \frac{\tilde{V}_{ji}}{2}\left(\vec{s}_{i} - \vec{s}_{j}\right)^2,
\end{align}
where the potential $V_{ji}$ depends on the distance between magnetic solitons and intrinsic features, such as chirality and core size. Here, we take up to lowest non-zero order in the vector distance, such that $\tilde{V}_{ji}$ is a constant that incorporates the dependence on the intrinsic features. On the other hand, the non conservative force due to electron polarization and scattering with magnetization on each layer can be written as:
\begin{align}\label{eq:NonConserv}
F_{i\, a}=-\sigma_i \mathcal{A}_{ij _a} -\sigma_i \Delta s_{a'} \mathcal{B}_{ij _{a',a}}.
\end{align}
Here, the sum over repeated index $a'$ is assumed,  we use a Taylor expansion in $\vec{m}_j$, and $\Delta s_a$ is the $a$-component of the vector distance $\Delta\vec{s}_{ij}=\vec{s}_j -\vec{s}_i$ between the solitons. Besides, we define matrices $\mathcal{A}_{ij}, \mathcal{B}_{ij}$, with elements:
\begin{align}\label{eq:AB_NonConserv}
\mathcal{A}_{ij_a} &=\int_{V_i} d^3r_i \,  (\vec{m}_i \times \vec{m}_j) \cdot \partial_a \vec{m}_i,\nonumber\\
\mathcal{B}_{ij_{a',a}} &= \int_{V_i} d^3r_i \,  \vec{m}_i \cdot (\partial_{a'}\vec{m}_j \times \partial_{a}\vec{m}_i),
\end{align}
where the integrals are performed in the volume space of the $i$-layer. Note that $\vec{r}_i$ is the vector position in the space volume of the $i$-th layer while $\vec{s}_i$ is the magnetic soliton position in the $i$-th layer. The gyrovector, dissipation dyadic, conservative and non conservative forces depend on the magnetization texture hosted in the layers. In general, the tensors $\mathcal{A}_{ij_a}$ and $\mathcal{B}_{ij_{a',a}}$ depend largely on the intrinsic features.
To calculate the forces we use the 
Image Vortex Ansatz (IVA) \cite{mertens2000dynamics},  which corresponds to an edge side free model, in agreement with the simulations.
Assuming that the magnetization is described as 
$\vec{m}_{i} = \Pi_i\cos\theta\hat{z} + c_{i}\,\sin\theta\,\left(-\sin\varphi_{i}\, \hat{x} + \cos\varphi_{i}\, \hat{y}\right)$, the IVA is given by:
\begin{subequations}
\begin{align}
    \cos\theta &= \sqrt{1 -\left( \frac{2r_c r}{r_c^2 + r^2}\right)^2}, \quad r \leq r_c, \\
    \cos\theta &= 0, \quad r > r_c, \\
    \varphi_{i} &= \arctan\left( \frac{y_{i} - s_{i} \sin\phi_{i}}{x_{i} - s_{i} \cos\phi_{i}} \right) 
    + \arctan\left( \frac{y_{i} - \frac{\mathcal{R}_{i}^2}{s_{i}} \sin\phi_{i}}{x_{i} - \frac{\mathcal{R}_{i}^2}{s_{i}} \cos\phi_{i}} \right) 
    - \phi_{i}.
\end{align}
\end{subequations}
Here, $(x_{i}, y_{i})$ represent the coordinates relative to the disk center, $r$ represents the radial coordinate, $r_c$ represents the vortex core size, $\phi_{i}$ denotes the polar angle, and $s_{i}$ is the radial distance of the vortex core. Additionally, $\mathcal{R}_{i}$ corresponds to the characteristic radius of the disk.

When the vortices have the same chirality, i.e., $c_i =c_j$, the conservative and non-conservative forces over each vortex are given by (in polar coordinates):
\begin{align}
-\frac{\partial E_i}{\partial s_i} &= -(k_i+\Gamma_i)s_i + \kappa_i (s_i-s_j \cos(\phi_i - \phi_j))\nonumber\\
-\frac{1}{s_i}\frac{\partial E_i}{\partial \phi_i} &= \kappa_i s_j\sin(\phi_i - \phi_j )
\label{cons_forces_coupled_vortex}
\end{align}
and \eqref{eq:NonConserv} can be explicitly written as
\begin{align}
F_{s_i} &= -\sigma_i B_i s_j \sin(\phi_i -\phi_j) \nonumber\\
F_{\phi_i} &= -\sigma_i B_i (-s_i+s_j \cos(\phi_i -\phi_j)).
\label{noncons_forces_coupled_vortex}
\end{align}
here $\Pi_i$ is the polarization direction of VC in the $i$-th layer. In Eq.\eqref{cons_forces_coupled_vortex}, the term proportional to $k_i$ comes from the exchange and dipolar interaction within the layer, $\Gamma_i$ comes from the Oersted field interaction, and the term proportional to $\kappa_i$ includes the dipolar interaction between the layers. The parameters are given as:
\begin{align}
G_{i} &= -2\pi \Pi_{i}\Lambda_{i}, \nonumber\\
D_{\alpha\,i} &= \alpha_i\, \pi \Lambda_{i}\ln\left(\mathcal{R}/l_{ex\, i}\right), \nonumber\\
\Gamma_i &= 1.926  \gamma_0 J \Lambda_i \mathcal{R}_i c_i/2, \nonumber\\
B_i &= -2\pi \Lambda_i (2\Pi_i + \Pi_j)/3, \nonumber\\
k_i &= 2\pi \gamma_0 M_{s\, i} \Lambda_i \left( \frac{l_{ex}^2}{\mathcal{R}_i^2} \frac{1}{1-s_i^2/\mathcal{R}_i^2} + \frac{\Lambda_i}{\mathcal{R}_i} \zeta_i \frac{1}{1-s_i^2/(4\mathcal{R}_i^2)}\right), \nonumber \\
\kappa_i &=\frac{\gamma_0 \tilde V_{ji}}{\mu_0 M_{s\, i}}.
\end{align}
The first term in $k_i$ is related to the exchange interaction and the second term comes from the magnetostatic interactions within the layer (self-dipolar interactions) \cite{gaididei2010magnetic}.
The constants that define both the conservative and the non-conservative forces are intricately linked to fundamental properties, such as the size and charge of the solitons. These properties, in turn, can be finely tuned by external perturbations such as magnetic fields, electric fields, spin-polarized currents, and temperature variations. In this study, we focuse on the influence of the applied magnetic fields on these intrinsic properties.
The magnetic field is responsible for a change in the dimensions of the VC up to a critical field strength, where the potential to induce a reversal of the vortex polarization appears. Below this critical field threshold, it causes an increase in the size of the cores of vortices aligned with the orientation of the magnetic field, while simultaneously causing a decrease in the core dimensions of vortices aligned against the orientation of the magnetic field. Hence, in the first-order approximation, for small magnetic fields $\vec{H}$, we introduce a linear modulation of the constants $B_{i}$ and $\tilde{V}_{ij}$, depending on the influence of the magnetic field,
\begin{subequations}
    \begin{align}
        B_{i} \rightarrow B_{i}\left(1 + \xi_{i}\Pi_{i}H_{z}\right),\\
        \tilde{V}_{ij} \rightarrow \tilde{V}_{ij}\left(1 + \xi_{ij}\Pi_{i}\Pi_{j}H_{z}\right).
    \end{align}
\end{subequations}

By fitting the parameters $\xi_{i}$ and $\xi_{ij}$ locally for small variations, one obtains the proper dependence of the constants of the applied magnetic field.
Studying the precise values of the parameters $\xi_{i}$ and $\xi_{ij}$ is beyond the scope of this work. In the analytical calculations and in Fig.\ref{fig:2}, for simplicity, we considered that $\xi_{ij} = 0$, and we considered the variation of $B_{i}$ given as $B_{i}\rightarrow(1 \pm \xi)B_{i}$, where $+$ corresponds to the thin layer and $-$ to the thick layer. Here $\xi$ is the parameter associated with the tuning of the non-conservative force.

In the numerical simulations of the analytical model \eqref{eq:EOMpolar}, the nonlinear response of the coupled vortices has been observed for large variations of the parameter $\xi$. However, the introduction of higher order terms in the expansion of the effective terms enhances the nonlinearity of the response and allows to align the results more closely with experimental observations. For example, considering the Taylor expansion of the dyadic dissipation $D_{\alpha i}$ beyond the zeroth order has been crucial for observing nonlinear behavior in single vortex \cite{guslienko2011spin}.

In the expansion of the dissipation dyadic $D_{\alpha i}(\vec{X}_{i}) = D_{\alpha i}^{(0)} + D_{\alpha i}^{(2)} s_i^2$, the zero-order term, $D_{\alpha i}^{(0)}$, corresponds to the translation of the vortex as a rigid body. The second order term, $D_{\alpha i}^{(2)}$, accounts for the deformations of the VC as it is displaced from the disk center and is given by
\begin{subequations}
 \begin{align}
D_{\alpha i}^{(2)}= \alpha_i\frac{\pi \Lambda_i }{2\,l_{ex\, i}^2} \left[ 1-\left(\frac{l_{ex\, i}}{\mathcal{R}_i} \right)^4\right]
 \end{align}
\end{subequations}

\begin{figure}[!ht]
\centering
  \includegraphics[width=11cm]{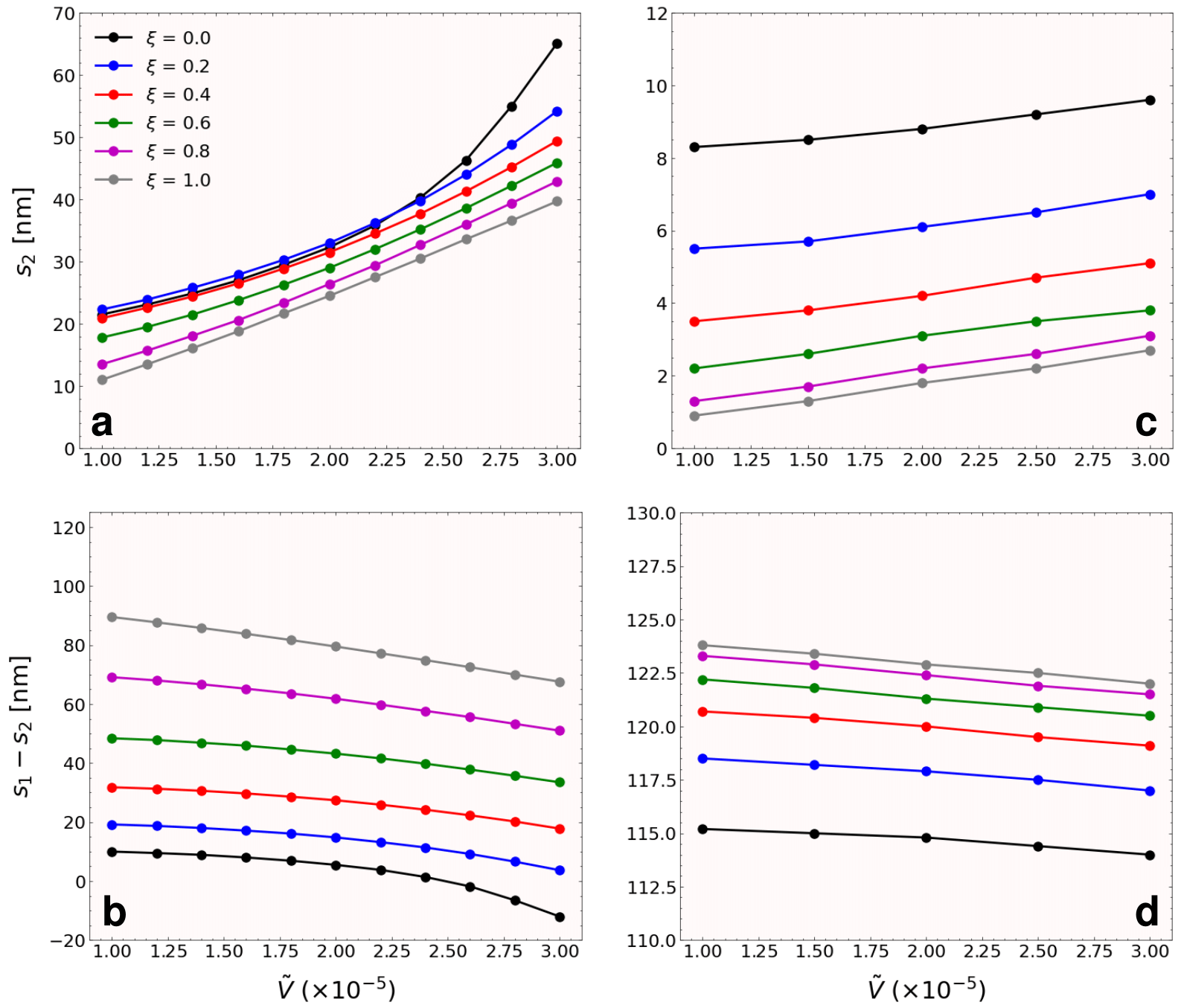}
  \caption{\textbf{VC displacement including higher order nonlinear terms:} (a) and (c) show the mean radial coordinate $s_2$ of the vortex in layer 2, while (b) and (d) show the distance between orbits $s_1-s_2$, of the vortex in layer 1 and 2. Different values of the dipolar interaction parameter $\tilde{V}$ and the control parameter $\xi$ were considered for a fixed applied current of 20 mA. In (a) and (b) the second order term $D_{\alpha i}^{(2)}$ was included, while in (c) and (d) this second order term was neglected.  
}
\label{sup_info_method}
\end{figure}

Fig. \ref{sup_info_method} shows a comparison of the average radial displacement of the VCs when considering only the zero-order term versus including the second-order term of the dyadic dissipation. Both variations of the dipolar interaction parameter $\tilde{V}$ and the parameter $\xi$ have been considered for a current of 20 mA. It is evident that the radial distances become significantly sensitive to variations of the parameters when the second order term is included, demonstrating that the system exhibits a much more nonlinear response. Therefore, fitting the control parameters to match the experimental results requires a detailed study of these parameters. However, as shown in the manuscript, this higher order expansion is not necessary to predict the nonlinear response of the coupled VCs.

\section*{Acknowledgements}
\textbf{Funding:}   This work was supported by the European Research Council within the Starting Grant No. 101042439 "CoSpiN"; the Deutsche Forschungsgemeinschaft (DFG, German Research Foundation) - TRR 173 - 268565370 (project B01); D.R.R. acknowledges funding from D.M. 10/08/2021 n. 1062 (PON Ricerca e Innovazione) funded by the Italian MUR; D.R.R. and G.F acknowledge funding from projects PRIN 2020LWPKH7 “The Italian factory of micromagnetic modelling and spintronics”, PRIN 20222N9A73 “SKYrmi-on-based magnetic tunnel junction to design a temperature SENSor – SkySens" funded by the Italian MUR and project number 101070287—SWAN-on-chip—HORIZON-CL4-2021-DIGITAL EMERGING-01. A.R. acknowledges support from CIP2022036\\
\textbf{Author contributions:} 
A.H. and G.D.L. performed and analyzed the  measurements with support from O.K., A.H. and A.K. carried out micromagnetic simulations  with support from P.P..\\
D.R.R., A.R. and G.F. developed the analytical theory and  performed the theoretical calculations.  A.H., P.P. and V.L. discussed the numerical simulations. A.H. and  P.P.  led this project.
A.H., A.K. and D.R.R. wrote the manuscript with the help of all the coauthors. All authors contributed to the scientific discussion and commented on the manuscript.\\
\textbf{Competing interests:} The authors declare that they have no competing interests.\\
\textbf{Data and materials availability: All data necessary to evaluate the conclusions of this paper are available upon reasonable request.}\\

\bibliography{sn-bibliography}

\end{document}